\documentclass[12pt]{aastex}
\usepackage{emulateapj5}
\usepackage{graphicx,psfig}
\def\ltsima{$\; \buildrel < \over \sim \;$}
\def\lsim{\lower.5ex\hbox{\ltsima}}
\def\gtsima{$\; \buildrel > \over \sim \;$}
\def\gsim{\lower.5ex\hbox{\gtsima}}

\newcommand{\be}{\begin{equation}}
\newcommand{\en}{\end{equation}}
\newcommand{\ergs}{\rm \ erg \; s^{-1}}

\def\cmdue {\rm \ cm^{-2}}

\def\msole {~M_{\odot}}

\begin{document}

\received{~~} \accepted{~~}
\journalid{}{}
\articleid{}{}

\title{The variable quiescence of Cen X-4}

\author{S.~Campana\altaffilmark{1}, G.L. Israel\altaffilmark{2},
L. Stella\altaffilmark{2}, F. Gastaldello\altaffilmark{3},
S. Mereghetti\altaffilmark{3}} 

\altaffiltext{1}{INAF-Osservatorio Astronomico di Brera, Via Bianchi 46, I--23807
Merate (Lc), Italy}

\altaffiltext{2}{INAF-Osservatorio Astronomico di Roma,
Via Frascati 33, I--00040 Monteporzio Catone (Roma), Italy}

\altaffiltext{3}{CNR-IASF, Istituto di Astrofisica Spaziale e Fisica Cosmica
Sezione di Milano ``G.Occhialini'', Via Bassini 15, I--20133 Milano, Italy}


\begin{abstract}
Cen X-4 is one of the best studied low-mass neutron star transients in
quiescence. Thanks to XMM-Newton large throughput, Cen X-4 was observed at the
highest signal to noise ever. This allowed us to disclose rapid ($>100$ s), 
large ($45\pm7$ rms in the $10^{-4}-1$ Hz range) intensity variability,
especially at low energies. In order to
highlight the cause of this variability, we divided the data into 
intensity intervals and fit the resulting spectra with the 
canonical model for neutron star transients in quiescence, i.e. an absorbed
power law plus a neutron star atmosphere. The fit is consistent with a
variable column density plus variability in (at least) one of the spectral
models. Variations in the neutron star atmosphere might suggest that accretion
onto the neutron star surface is occuring in quiescence; variations in the
power law tail should support the view of an active millisecond radio pulsar
emitting X-rays at the shock between a radio pulsar wind and inflowing matter
from the companion star. 
\end{abstract}

\section{Introduction}

Many Low Mass X--ray Binaries (LMXRB) accrete matter at very 
high rates, and therefore shine as bright X--ray sources, only sporadically. 
Among these systems are Soft X--ray Transients (SXRTs) hosting an
old neutron star (for a review see Campana et al. 1998a).
These systems alternate periods (weeks to months) of high X--ray luminosity,
during which they share the same properties of persistent LMXRBs,
to long (1--few 10 years) intervals of quiescence in which the X--ray luminosity
drops by up to 5--6 orders of magnitude. 

Cen X-4 is one of the best studied SXRTs. Bright X--ray
outbursts were detected in 1969 and 1979; the source remained quiescent since. 
During the 1979 outburst Cen X-4 reached a peak flux $L_X\sim 4\times
10^{37}\ergs$ (for a distance of $d \sim 1.2$~kpc, Kaluzienski, Holt \& Swank
1980). Type I bursts were observed, testifying to the presence of an accreting 
neutron star. 
Cen X-4 was observed several times in quiescence (Asai et al. 1996, 1998;
Campana et al. 1997, 2000; Rutledge et al. 2001). The spectrum was  
fit with a soft thermal component (neutron star atmosphere or black body) plus
a hard power law with photon index in the 1--2 range.  
Rutledge et al. (2001) reported that the 0.5--10 keV luminosity decreased by
$40\pm8\%$ in the 5 yr between the ASCA and Chandra observations. This
variability can be attributed to the power-law component; on the contrary
temperature variations in the thermal component are limited to $\lsim 10\%$. 
Short time variability was discovered by Campana et al. (1997)
during a ROSAT HRI pointing, with the source flux fading by a factor of $\sim 3$ 
on a timescale of $\lsim 4$ days.

Extensive spectroscopic and photometric measurements of the optical
counterpart in quiescence (V=18.7 mag) led to the determination of
the orbital period (15.1 hr; Chevalier et al. 1989; McClintock \& Remillard
1990; Cowley et al. 1988) and mass function ($\sim 0.2 \msole$, converting 
to a neutron star mass between $0.5-2.3 \msole$). 
The optical spectrum shows the characteristics of a
K5-7 main sequence star, contaminated by lines (e.g. H$\alpha$, H$\beta$ and
H$\gamma$) and continuum emission probably resulting from an accretion disk
(Shahbaz, Naylor \& Charles 1993). The latter was estimated to contribute
$\sim 80\%$, $\sim 30\%$, $\sim 25\%$ and $\sim 10\%$ of the quiescent optical
flux in the B, V, R and I bands, respectively.
A short wavelength HST/FOS spectrum of the quiescent optical counterpart
yielded a 1350--2200\,\AA\ luminosity of $0.6-4\times 10^{31}\ergs$.
These results were confirmed by higher quality HST/STIS spectra
(McClintock \& Remillard 2000). In particular, the UV spectrum appears to lie
on the extrapolation of the power law component seen in X rays.

In this paper we take advantage of an XMM-Newton observation of Cen X-4 to
investigate, in much higher detail its quiescent state. 
We describe the observation characteristics and data filtering in Section 2.
In Section 3 we discuss on spectral and timing results.
The interpretation of the data and discussion is presented in Section 4.
Our conclusions are in Section 5.

\section{Observation and data analysis}

The XMM-Newton Observatory (Jansen et al. 2001) comprises three $\sim 1500$
cm$^2$ effective area X--ray telescopes each with a European Photon Imaging
Camera (EPIC, 0.1--15 keV) at the focus. Two of the EPIC imaging spectrometers
use MOS CCDs (Turner et al. 2001) and one uses pn CCDs (Str\"uder et
al. 2001). Reflection Grating Spectrometers (RGS, 0.35--2.5 keV, den Herder et
al. 2001) are located behind two of the telescopes. 

Cen X-4 was observed on 2001 August 20--21. The EPIC cameras were operated in
the Prime Full Window mode with thin filters. We used the data generated by
the Pipeline Processing Subsystem in September 2001.
The observation is plagued by high background intervals due to soft proton
flares. 
We excluded them using an intensity filter: all events accumulated in a time
interval such that the rate, excluded the source region, exceeded 0.3 c
s$^{-1}$ in the 10--12 keV band for each of the two MOS cameras and 0.8 c  s$^{-1}$
for the pn camera were rejected. We adopted these thresholds, which are
somewhat higher than the standard ones, motivated by the strength of
our source ($\sim 1$ c s$^{-1}$ in the 0.2--8 keV band of the pn, see below). 
We verified  
that results obtained by using standard thresholds are consistent with the
cleaner (but shorter) exposures. 
We obtained net (original) exposure times of 32 (52) ks, 32 (52) ks, 24 (40) ks
for MOS1, MOS2 and pn, respectively.    
These represent the deepest observation of a neutron star SXRT ever carried
out. RGS spectra were heavily affected by soft proton flares due to the dispersion
over the field of view. Optical monitor was not operated.

For the pn, single and double pixel events were selected
(patterns 0 to 4), for the MOS, events corresponding to patterns from 0 to 12
were selected. Source spectra were extracted from a circular region of $45''$ 
for the two MOS cameras and $40''$ for pn camera, corresponding to an encircle 
energy fraction of $\sim 90\%$. Background spectra have been 
extracted from nearby circular regions $2'$ and $2.5'$ for the MOS1 and MOS2,
respectively. For the pn we considered two circular regions of $40''$ and
$60''$ in the same CCD of the source.
Following the prescriptions of Snowden et al. (2002) we extracted the spectra
(including the FLAG=0 option) obtaining 9609 counts (0.29 c s$^{-1}$), 9829
counts (0.30 c s$^{-1}$), 28673 counts (1.1 c s$^{-1}$) from the MOS1, MOS2
and pn in the full band, respectively. Note that pile up was negligible ($<1\%$). 

Response matrices and ancillary region files were generated with the SAS
(v5.3.3) tasks {\tt rmfgen} and {\tt arfgen}. The spectral analysis was 
carried out in the 0.5--10 keV energy range for all the instruments using
XSPEC (v11.2). All spectral uncertainties are given at $90\%$ confidence level 
for one interesting parameter ($\Delta \chi^2=2.71$).

\section{Results}

\subsection{Spectral analysis: I}

We first consider the total spectrum. We rebinned the MOS and pn spectra so as to
have 25 and 30 counts per spectral channel, respectively. We verified that the
background spectra, especially at high energies, are not dominated by residual
soft proton flares. 
In particular, pn and MOS background spectra show the same
behaviour and decrease by a factor of $\sim 2$ from 2 to 10 keV as expected
for normal conditions (Lumb et al. 2002).  

\begin{figure*}
\begin{center}
\psfig{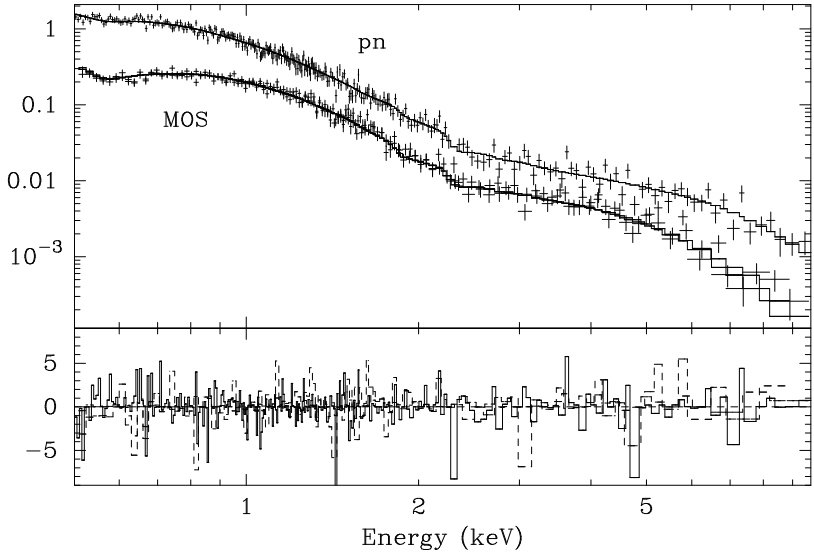}
\caption{MOS and pn spectrum of Cen X-4 in the 0.5--10 keV range with the best 
fit model (absorbed neutron star atmosphere plus power law) overlaid. In the
lower panel the differences between data and model in terms of $\chi^2$. Dashed
lines in the lower panel refer to MOS data.}
\label{spe}
\end{center}
\end{figure*}

The spectrum cannot be fit with any single component model. We then tried the
canonical spectrum for SXRT in quiescence (e.g. Campana et al. 1998b; Campana
2001), i.e. an absorbed neutron star hydrogen atmosphere plus a power law (we
adopted the absorption model {\tt TBABS} by Wilms, Allen \& McCray 2000 and
the neutron star atmosphere model by G\"ansicke, Braje \& Romani 2002). 
We included also a constant factor to account for the
mismatch between the different EPIC instruments (Kirsch 2002). 
This model provides a reduced $\chi^2_{\rm red}=1.07$ for 518 degrees of
freedom (d.o.f.), corresponding to a null hypothesis probability (n.h.p.) of
$12\%$ (see Fig. \ref{spe}). Neutron star atmosphere models with iron or solar 
composition or a disk black body model provided worse fits with $\chi^2_{\rm
red}>1.3$. A black body model formally provided a better fit to the data but
with a null column density ($<1.4\times 10^{20}\cmdue$ see Table
\ref{spetot}). Taking instead a column density of $9\times 10^{20}\cmdue$ (in
line with expectations from optical data) we obtained a worse fit
($\chi^2_{\rm red}=1.18$). 

The unabsorbed 0.5--10 keV flux of the power law plus neutron star atmosphere
models is $2.3\times 10^{-12}\ergs\cmdue$ that at a distance of 1.2 kpc 
translates into a luminosity of $3.9\times10^{32}\ergs$.
In the 0.5--10 keV band the atmosphere model accounts for $63\%$ of the total 
luminosity.  

\subsection{Temporal analysis}

\begin{figure*}
\begin{center}
\psfig{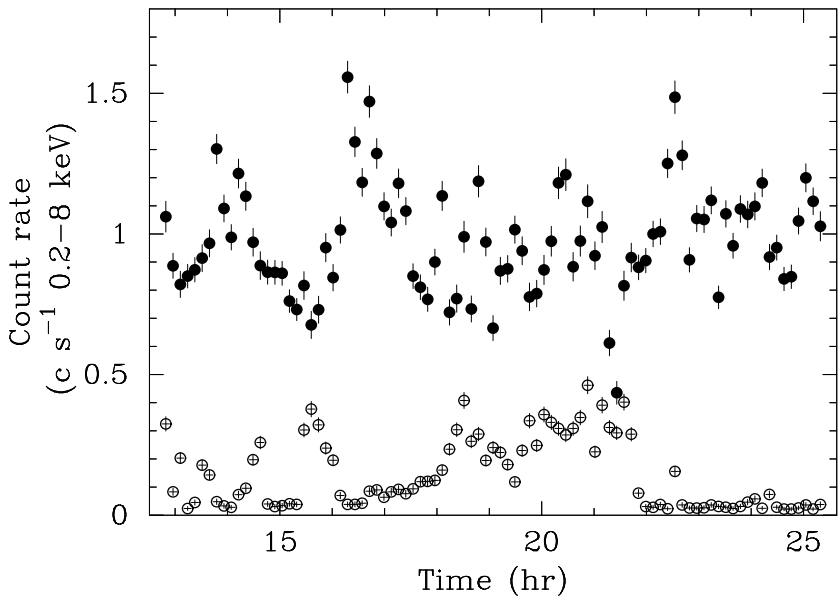}
\caption{Background subtracted 0.2--8 keV pn light curve of Cen X-4. The bin
size is 500 s. Open dots represent the background light curve scaled to the
source extraction area (which has been already subtracted in the light curve above).} 
\label{curve}
\end{center}
\end{figure*}

For timing analysis we considered the entire observation, 
including high background time intervals. Besides source light curves, we
extracted background light curves which we subtract (including the error on
the background rate) from the source light curve. 
The resulting light curve is clearly not constant (Fig. \ref{curve}). 
The importance of the background is in any case minimal apart from the 
minimum around 21.5 hr in Fig. \ref{curve} when the background reached $40\%$ of
the Cen X-4 rate.

\begin{figure*}
\begin{center}
\psfig{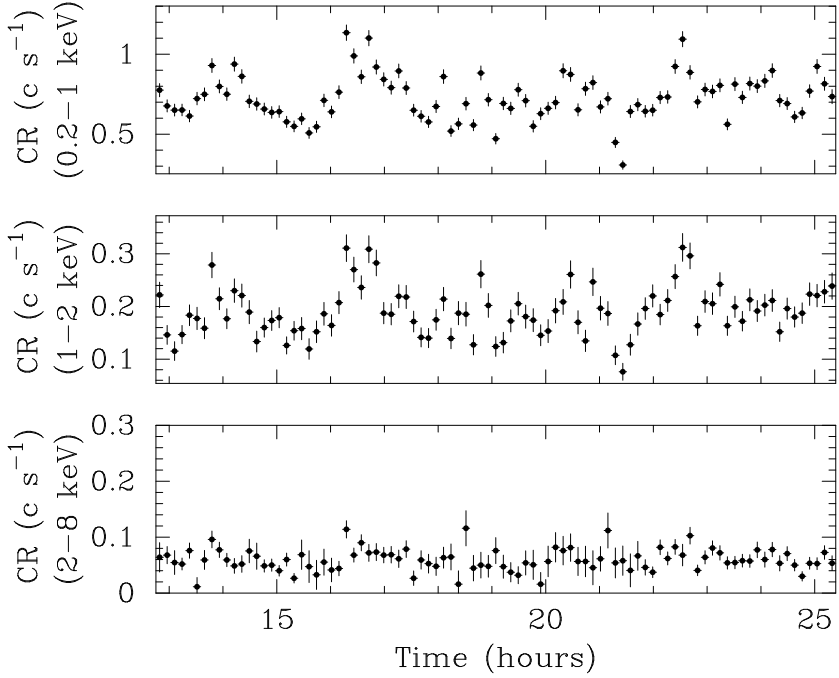}
\caption{Background subtracted 0.2--8 keV pn light curve of Cen X-4 in three
energy bands: 0.2--1 keV (upper panel); 1--2 keV (middle panel) and 2--8 keV
(lower panel). The bin size is 500 s.}
\label{3curve}
\end{center}
\end{figure*}

We extracted light curves in selected energy bands: 0.2--1 keV, 1--2 keV
and 2--8 keV (see Fig. \ref{3curve}). A smaller amplitude of the variability
at high energies is apparent. Taking a bin time of 500 s (in order to have
a mean of 30 counts per bin in the 2--8 keV light curve) we fit the
light curves with a constant and we obtained
$\chi^2_{\rm red}=13.7,\ 4.8,\ 1.6$ in the three energy bands, respectively.
Possible flare-like events can be identified around 14 hr, 16.5 hr and 22.5 hr
in Fig. \ref{curve}. Rise times are about 20 min. A dip-like feature is also
visible around 21.5 hr. 

We obtained a power spectrum of the entire observation by using the cleanest pn
exposure. The power spectrum shows a strong noise component below $\sim
0.01$ Hz (i.e. a timescale longer than $\sim 100$ s), consistent with the
variability discussed above. The rms variability amounts to $45\pm7\%$ in the
$10^{-4}-1$ Hz band. The power spectrum can be well fit with a power law with
index $-1.2\pm0.1$. No significant periodicities or quasi-periodic
oscillations are seen. The power-law index of the X--ray power spectrum
is consistent with the optical one which is characterized by an index of $\sim
-1$, $-1.5$ (Hynes et al. 2002; Zurita et al. 2003).
 
\subsection{Spectral analysis: II}

\begin{figure*}
\begin{center}
\psfig{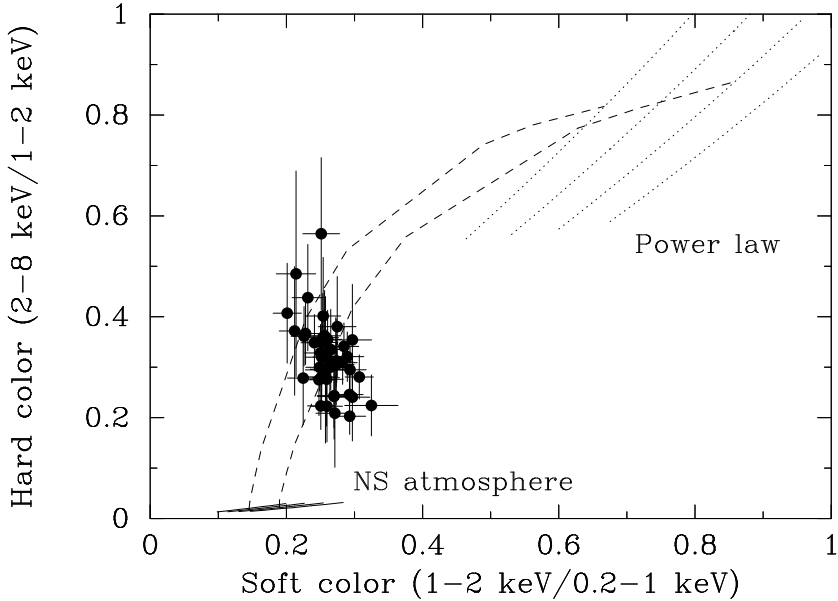}
\caption{Color-color diagram of the pn background subtracted light curve. The
soft color is defined as the ratio of the counts in the 0.2--1 keV to 1--2 keV,
the hard color as the ratio of the counts in the 2--8 keV to 1--2 keV. On top
of this we depict hardness ratios for an absorbed power law model (with photon 
index from 1.2 to 2) and an absorbed neutron star atmosphere model (with
temperatures from 70 to 100 eV) for different values of column densities (3,
5, 7, $9\times 10^{20}\cmdue$, dotted lines). Two different (dashed) curves
connect the single component models with increasing contribution from the two
for column densities of 5 and $9\times 10^{20}\cmdue$.}  
\label{color}
\end{center}
\end{figure*}

To study possible spectral changes related to the variability we constructed
the color-color diagram made by a soft color (1--2 keV / 0.2--1
keV) versus a hard color (2--8 keV / 1--2 keV, see Fig. \ref{color}).
Spectral variations are visible in the data.
We plot in the same diagram the colors expected for single (absorbed) power
law and neutron star atmosphere spectra for a set of parameters. 
As expected from the spectral analysis (section 3.1) these single component
models cannot account for the data.
In addition, we plot two lines with different column densities and different
fractions of the two components models. We note that variations appear to be
consistent with variations in the column density across the observation.
Alternatively a variation in the fraction of the power law flux to neutron star
atmosphere flux can account for the observed variation.

To check these hypotheses we carry out a more detailed spectral analysis. We
concentrate on pn data, since these provide a factor of $\sim 2$ more counts
than the two MOS cameras together. We divided the data in the pn
light curve (1,000 s bin) at three different count rates: above 1.3 c
s$^{-1}$, between 1.0 and 1.3 c s$^{-1}$ and below 1.0 c s$^{-1}$.  We then
fit the three corresponding spectra in order to search for possible
differences.  
We consider the same absorbed power law plus neutron star atmosphere
model. The same spectrum for the three count rate spectra is
clearly not acceptable ($\chi^2_{\rm red}=2.60$). Letting the column density
vary freely produces an improvement in the fit but not enough to make the fit acceptable
($\chi^2_{\rm red}=1.51$; the column density is smaller at larger count
rates). This shows that the variability cannot be fully ascribed to a column
density variation. 
Better results can be obtained allowing (at least) one of the spectral
component to vary
together with the column density\footnote{In the case of a free neutron star atmosphere
component, we consider only a variation in temperature and check that indeed
the neutron star radius remains stable.} (Table \ref{highbase}).  
With an additional parameter we obtained better results, however we cannot
decide on a statistical basis which of the two fits is better. No clear
correlations are observable. The column density is within the errors for both
these new fits. 
In the case of a variable neutron star atmosphere, the temperature 
increases with the count
rate, in the case of a variable power law, the power law photon index steepens
as the count rate increases.

\section{Discussion}

The quiescent state of Cen X-4 has been recognised to be variable both
on long times scales ($\sim 40\%$ in 5 yr, Rutledge et al. 2001) and on
shorter timescales (factor of $\sim 3$ in a few days, Campana et al. 1997).   
During this XMM-Newton observation X--ray variability has been observed (at a
level of $\sim 45\%$ rms) on even shorter timescale (down to $\sim 100$ s) thanks
to the large collecting area. Variability on such a short timescale would have
been missed if observed with any previous X--ray satellite. 

In the pn light curve three flare-like events can be identified. 
X--ray flares have been observed in the transient black hole candidate V404
Cyg also in quiescence (Wagner et al. 1994; Kong et al. 2002).  
Flare activity has been recently reported also in the optical for a number of
transient black holes during quiescence as well as for Cen X-4 
(Zurita, Casares \& Shabhaz 2003;
Hynes et al. 2002). Flares occur on timescales of minutes to a few hours, with
no dependence on orbital phase. R-band luminosities are in the range 
$10^{31}-10^{33}\ergs$. The mean duration of optical flares in Cen X-4 is 21
min. This is similar to what observed in the X--rays.

Several mechanisms have been proposed for optical flares (Zurita et al. 2003;
Hynes et al. 2002). Chromospheric activity, instabilities in the mass transfer
rate from the companion and viscous instabilities in an accretion disk have been
ruled out for different reasons (Zurita et al. 2003). 
Reprocessing of X--ray variations in the outer disk regions has also
been considered less likely due to the large ratio between optical and X--ray
emission (Zurita et al. 2003). In passing, we note that in the X--ray light
curve three flare events can be identified (see Fig. \ref{curve}). The
durations and energetics of the first two flares are comparable: 0.9 and 1.2
hr and 1.8 and $4.4\times 10^{35}$ erg, respectively. The last flare is much
shorter and less energetic (0.1 hr and $0.6\times 10^{35}$ erg).

Long term X--ray variability in the quiescent state of Aql X-1 has also been
reported based on Chandra data. In particular, Rutledge et al. (2002) showed
that these variations cannot be caused by variations in the power law only,
but can be accounted for by a varying temperature of the neutron star 
atmosphere model. These variations are not monotonic but show
first a decrease, then an increase and finally a decrease in time. 
This would rule out the deep crustal heating mechanism for SXRTs in quiescence.
Campana \& Stella (2003) provided a different interpretation of these data
(plus an additional BeppoSAX observations): a varying column density and 
power law component together with a stable neutron star atmosphere model.

This latter model is motivated by the recent discovery of a millisecond radio 
pulsar PSR J1740--5340 in the globular cluster NGC 6397 (D'Amico et al. 2001; 
Ferrario et al. 2001).
This pulsar shows irregular eclipses in radio and it is also emitting in 
X--rays (Grindlay et al. 2001). The X--ray emission is likely powered 
by the interaction of the relativistic wind of the radio pulsar with matter 
outflowing from the companion. This shock emission mechanism has also
been put forward to explain the X--ray emission of the radio pulsar 
PSR 1259--63 orbiting a Be companion (Tavani \& Arons 1997).
This model is not disproved by the data. The power law photon index is between 
1.5 and 2 as one would expect from simple modelling based synchrotron
emission. The luminosity increases as the power law photon index increases
as observed in PSR 1259--63 for $\Gamma <1.8$. 

\section{Conclusions}

We observed large ($\sim 45\%$) r.m.s. X--ray variability in the quiescent
state of Cen X-4. 
Small spectral variations are observed as well.
These can be mainly accounted for by a variation in the column density together 
with another spectral parameter. Based on the available spectra we cannot
prefer a variation of the power law versus a variation in the temperature of
the atmosphere component (even if the first is slightly better in terms of
reduced $\chi^2$). This variability can be accounted for by accretion
onto the neutron star surface (e.g. Rutledge et al. 2002) or by the variable
interaction between the pulsar relativistic wind and matter outflowing from
the companion in a shock front (e.g. Campana \& Stella 2003). 


\begin{acknowledgements}
We thank M. Guainazzi for his advice in the data analysis.  
\end{acknowledgements}

\newpage

\begin{table*}[!htb]
\begin{center}
\caption{\label{spetot} Spectral models for the entire XMM-Newton observation.}
\begin{tabular}{cccccc}
\hline
Model    & $N_H$               &$k\,T$    &Photon       & $\chi^2_{\rm red}$& n.h.p.  \\ 
         &($10^{20}$ cm$^{-2}$)& (eV)     &index        &    (d.o.f.)       &         \\
\hline
Hydrogen & $8.3\pm3.5$         &$85\pm3$  &$1.56\pm0.09$& 1.07 (518)        & $12.0\%$ \\
Solar    & $11.1\pm1.5$        &$>14$     &$3.41\pm0.08$& 2.88 (518)        &\ $0.0\%$ \\
Iron     & $4.5\pm1.0$         &$<273$    &$2.79\pm0.09$& 2.23 (518)        &\ $0.0\%$ \\
Disk BB  & $7.3\pm2.2$         &$219\pm13$&$1.73\pm0.13$& 1.31 (518)        &\ $0.0\%$ \\
Blackbody& $<1.4$             &$186\pm5$ &$1.64\pm0.09$& 1.06 (518)        & $16.1\%$ \\
\hline
\end{tabular}
\end{center}
\end{table*}

\begin{table*}[!htb]
\begin{center}
\caption{\label{highbase} Standard spectral models fits for the high and low
count rate interval.}
\begin{tabular}{cccccc}
\hline\hline
Model    & $N_H$               &$k\,T$   &Photon index         & $\chi^2_{\rm red}$& N.h.p.  \\ 
         &($10^{20}$ cm$^{-2}$)& (eV)    &(hard component)     &     (d.o.f.)         &         \\
\hline
Low      & $6.0\pm2.9$       &$87\pm1$ &$1.53\pm0.17\ (40\%)$& 2.60 (450)        &\ $0\%$ \\
Medium   &                   &         &                     &                   &        \\
High     &                   &         &                     &                   &        \\
\hline
Low      & $15.1\pm3.6$      &$78\pm9$ &$1.60\pm0.14\ (35\%)$& 1.51 (448)        &\ $0\%$ \\
Medium   & $8.9\pm4.3$       &         &                     &                   &        \\
High     & $4.4\pm3.4$       &         &                     &                   &        \\
\hline
Low      & $5.1\pm2.6$       &$86\pm7$ &$1.47\pm0.13 \ (46\%)$& 1.02 (446)        & $37\%$ \\
Medium   & $4.6\pm2.6$       &$91\pm5$ &$ \hskip 2.2cm (38\%)$&                   &        \\
High     & $6.0\pm2.1$       &$99\pm7$ &$ \hskip 2.2cm (30\%)$&                   &        \\
\hline
Low      & $9.6\pm4.6$      &$85\pm6$  &$1.32\pm0.19 \ (33\%)$& 1.03 (444)        & $34\%$ \\
Medium   & $5.4\pm3.7$      &          &$1.62\pm0.16 \ (42\%)$&                   &        \\
High     & $4.6\pm3.8$      &          &$2.11\pm0.20 \ (53\%)$&                   &        \\ 
\hline\hline
\end{tabular}
\end{center}
\tablecomments{
The emitting radius of the neutron star atmosphere component has been
constrained to be the same in all the three observations.
In the case of a varying neutron star atmosphere model the radius is
$9.7\pm2.4$ km (intrinsic radii), whereas for a varying 
power law we have a radius of $11.5\pm6.0$ km.} 
\end{table*}

\newpage

\begin{references}

\reference {}
Asai, K., et al. 1996, PASJ 48 257

\reference {}
Asai, K., et al. 1998, PASJ 50 611

\reference {}
Campana, S. 2001, in the proceedings of the Bologna conference on ``X--ray
Astronomy 1999: Stellar Endpoints, AGN, and the Diffuse X-ray Background",
eds. N.E. White, G. Malaguti, G.G.C. Palumbo, A.I.P. Conf. Proc. 599 63

\reference {}
Campana, S., Stella, S. 2003, ApJ submitted

\reference {}
Campana, S., et al. 1997, A\&A 324 941

\reference {}
Campana, S., et al. 1998a, A\&A Rev. 8 279

\reference {}
Campana, S., et al. 1998b, ApJ 499 L65

\reference {}
Campana, S., et al. 2000, A\&A 358 583

\reference {}
Chevalier, C., Ilovaisky, S.A., van Paradijs, J., Pedersen, H., van der Klis,
M. 1989, A\&A 210 114 

\reference {}
Cowley, A.P., Hutchings, J.B., Schmidtke, P.C., Hartwick, F.D.A.,
Crampton, D., Thompson, I.B. 1988, AJ 95 1231 

\reference {}
D'Amico, N., et al. 2001, ApJ 561 L89

\reference {}
den Herder, J.W., et al. 2001, A\&A 365 L7

\reference {}
Ferrario, F., Possenti, A., D'Amico, N., Sabbi, E. 2001, ApJ 561 L93

\reference {}
G\"ansicke, B.T., Braje, T.M., Romani, R.W. 2002, A\&A 386 1001

\reference {}
Grindlay, J., Heinke, C.O., Edmonds, P.D., Murray, S.S., Cool, A.M. 2001, ApJ
563 L53

\reference {}
Hynes, R. I., Zurita, C., Haswell, C. A., Casares, J., Charles, P. A.,
Pavlenko, E., Shugarov, S., \& Lott, D. A. 2002, MNRAS 330 1009

\reference {}
Jansen, F., et al. 2001, A\&A 365 L1

\reference {}
Kaluzienski L.J.,  Holt, S.S.,  Swank, J.H. 1980, ApJ 241 779

\reference {}
Kong, A.K.H., McClintock, J.E., Garcia, M.R., Murray, S.S., Barret, D. 2002
ApJ 570 277

\reference {}
Lumb, D.H., Warwick, R.S., Page, M., De Luca, A. 2002 A\&A 389 93

\reference {}
McClintock, J.E., Remillard, R.A. 1990, ApJ 350 386

\reference {}
McClintock, J.E., Remillard, R.A. 2000, ApJ 531 956

\reference {}
Nowak, M., Heinz, S., Begelman, M. 2002, ApJ 573 778

\reference {}
Rutledge, R.E., Bildsten, L., Brown, E.F., Pavlov, G.G., Zavlin, V.E.
2001, ApJ 551 921 

\reference {}
Rutledge, R.E., Bildsten, L., Brown, E.F., Pavlov, G.G., Zavlin, V.E., 2002,
ApJ 577 346

\reference {}
Shahbaz, T., Naylor, T., Charles, P.A. 1993, MNRAS 265 655

\reference {}
Tavani, M., Arons, J. 1997, ApJ 477 439

\reference {}
Turner, M. et al., 2001, A\&A 365 L27

\reference {}
Wagner, R.M., Starrfield, S.G., Hjellming, R.M., Howell, S.B., Kreidl, T.J.
1994, ApJ 429 L25

\reference {}
Wilms, J., Allen, A., McCray, R. 2000, ApJ 542 914

\reference {}
Zurita, C., Casares, J., Shahbaz, T. 2003, ApJ 582 369

\end{references}
\end{document}